\documentclass[epj]{svjour}
\usepackage{graphicx}
\usepackage{amssymb,amsfonts,amsmath}
\begin{document}

\title{Static and Dynamic Structure Factors with Account of the Ion Structure for High-temperature Alkali and Alkaline Earth Plasmas\thanks{\bf{Dedicated to the 100th birthday of N. N. Bogolyubov}}}

\author{S. P. Sadykova\inst{1}\thanks{\emph{e-mail:} saltanat@physik.hu-berlin.de}, W. Ebeling\inst{1}\thanks{\emph{e-mail:} ebeling@physik.hu-berlin.de} \and I. M. Tkachenko\inst{2}\thanks{\emph{e-mail:} imtk@mat.upv.es}
}                     
%
%
\institute{Institut f\"{u}r Physik, Humboldt Universitat zu Berlin, Newtonstr. 15, 12489 Berlin, Germany \and Instituto Universitario de Matem\'{a}tica Pura y Aplicada, Universidad Polit\'{e}cnica de Valencia, Camino de Vera s/n, 46022 Valencia, Spain}

\date{Received: date / Revised version: date}
%
\abstract{The $e-e$ , $e-i$, $i-i$ and charge-charge static structure factors are calculated for alkali and Be$^{2+}$ plasmas using the method described by Gregori et al. in \cite{bibGreg2006}. The dynamic structure factors for alkali plasmas are calculated using the method of moments \cite{bibAdam83}, \cite{bibAdam93}. In both methods the screened Hellmann-Gurskii-Krasko potential, obtained on the basis of Bogolyubov's method, has been used taking into account not only the quantum-mechanical effects but also the ion structure \cite{bib73}.
} 
\titlerunning{Stat. and Dyn. Struc. Factors with Account of the Ion Structure}
\authorrunning{S. P. Sadykova et al.}
\maketitle
\section{Introduction}
\label{intro}
\indent The structure and thermodynamic properties of alkali and alkaline earth plasmas are of basic interest and of importance for high-temperature technical applications. Near  the critical point and at higher temperatures the materials are in the thermodynamic state of strongly coupled plasmas. Here we will go far beyond the critical point to the region of nearly fully ionised plasmas which is $T\geq 30$  kK for alkali and $T\geq 100$ kK for alkaline earth plasmas. The investigation of thermodynamic properties of alkali plasmas under extreme conditions is not only important for basic research. There are many applications, e.g. in material sciences, geophysics and astrophysics. Furthermore, these studies throw some light on the complex picture of phase transitions in metal vapors which play an outstanding role in technological applications.\\
\indent Over the recent decades a considerable amount of effort has been concentrated on the experimental \cite{bib74}, \cite{bib75} and theoretical \cite{bib79}-\cite{bib82} investigation of the behavior of alkali metals in the liquid and plasma state expanded by heating toward the liquid-vapor critical point. High-temperatures alkali plasmas are widely applied in many technical projects. For instance, Li is an alkali metal of considerable technological interest. Lithium is planned to be used in inertial confinement fusion, solar power plants, electrochemical energy storage, magnetohydrodynamic power generators and in a lot of other applications. Recent advances in the field of extreme ultraviolet EUV lithography have revealed that laser-produced $Li$ plasmas are source candidates for next-generation microelectronics \cite{bib60}. For this reason we believe that the study of basic properties of alkali plasmas are of interest. In the previous work we studied Li$^+$ plasma \cite{bib73}. In this work, we consider Li$^+$, Na$^+$, K$^+$, Rb$^+$, Cs$^+$ and Be$^{2+}$ plasmas. For simplicity of the calculations we take into account here only single ionization for alkali plasmas ($n_e=n_i$) and doubled for beryllium plasma ($n_e=2n_i$), where $n_e$, $n_i$ are the concentrations of electrons and ions respectively. Li, Na, K, Rb, Cs atoms have one outer electron and Be$^{2+}$ has two outer electrons. In the table \ref{tab:1} the ionization energies of alkali atoms and beryllium atom are presented.
\begin{table}
\caption{The ionization energies $I_i$ $(eV)$ of alkali and Be atoms }
\label{tab:1}       
\begin{tabular*}{88mm}{llllllll}
\hline\noalign{\smallskip}
&H&Li&Na&K&Rb&Cs&Be  \\
\noalign{\smallskip}
\hline\noalign{\smallskip}
$I_1$&13.595&5.39&5.138&4.339&4.176&3.893&9.306 \\
$I_2$&-&75.62&47.29&31.81&27.5&25.1&18.187 \\
\noalign{\smallskip}\hline
\end{tabular*}
\end{table}
 Correspondingly we will study temperatures around $30000 K$ for alkali and $100000 K$ for Be$^{2+}$ plasmas, where most of outer electrons are ionized, but the rest core electrons are still tightly bound.\\
 \indent Recently, X-ray scattering has proven to be a powerful technique in measuring densities, temperatures and charge states in warm dense matter regimes \cite{bibRiley}. In inertial confinement fusion and laboratory astrophysics experiments the system demonstrates a variety of plasma regimes and of high interest are the highly coupled plasmas $\Gamma_{ii}\geq 1$ ($\Gamma_{ii}=z^2e^2/(4\varepsilon_0 k_BT r_{ii})$ with $r_{ii}=(3/4\pi n_i)^{1/3}$  being the average ion-ion distance, $e$ is the electric elementary charge and $z$ - the ionic charge) and the electron subsystem exhibiting partial degeneracy. Such regimes can be often found during plasma-to-solid phase transitions. Recent experiments with a solid density Be plasma have shown a high level of ion-ion interaction and their interpretation must account for significant strong coupling effects. The present study is devoted to the study of the static (SSF) and dynamic (DSF) structure factors for alkali and Be$^{2+}$ plasmas at temperatures $T\geq 30$ kK and $T\geq 100$ kK respectively. The structure factor (SF) is the fundamental quantity that describes the X-ray scattering plasma cross-section. Since the SF is related to the density fluctuations in the plasma, it directly enters into the expression for the total cross-section. In the case of a weakly coupled plasma SF can be obtained within the Debye-H{\"u}ckel theory or the random phase approximation (RPA), while at moderate coupling the RPA fails to predict the correct spatial correlations. However, recent works by Gregori et al. \cite{bibGreg2006}, \cite{bibGreg2007} and by Arkhipov et al. \cite{bibArkhipov2008} have shown that the technique developed in the classical work of Bogolyubov provides sufficiently reliable expressions of SF in moderately coupled plasmas.\\
\indent For the determination of static and dynamic structure factors one needs to have a screened pseudopotential as an essential input value. The semiclassical methods allow to include the quantum-mechanical effects by appropriate pseudopotentials which resolve the divergency problem at small distances. This method was pioneered by Kelbg, Dunn, Broyles, Deutsch and others \cite{{bib36}} - \cite{bib58}; later it was significantly improved in \cite{bib37}-\cite{bib29}. These models are valid for highly temperature plasmas when the ions are bare or there is no significant influence of the ion shell structure. In order to correctly describe alkali plasmas at moderate temperatures one needs to take into account the ion structure. For example, to describe the behaviour of alkali plasmas, the short range forces between the charged particles are of great importance. For alkali plasmas at small distances between the particles deviations from the Coulomb law are observed which are mainly due to the influence of the core electrons. The method of model pseudopotentials describing the ion structure was pioneered by Hellmann. Hellmann demonstrated, using the Thomas-Fermi model, that the Pauli exclusion principle for the valence electrons can be replaced by a non-classical repulsive potential \cite{bib72}. This method was later rediscovered and further developed for metals by Heine, Abarenkov and Animalu \cite{bib46}, \cite{bib47}. Heine and Abarenkov proposed a model, where one considers two types of interaction: outside of the shell, where the interaction potential is purely Coulomb one and inside, where it is constant. Parameters of these model potentials were determined using the spectroscopic data. Later on different pseudopotential models were proposed. For a more detailed review we refer to \cite{bib47}. All these models are characterized by one disadvantage. Their Fourier transforms (formfactors) are not sufficiently convergent when the Fourier space coordinate goes to infinity. Gurskii and Krasko \cite{bib51} proposed a model potential which eliminates this problem and provides smoothness of the pseuopotential inside the shell giving its finite value at small distances. The first attempt to construct the model pseudopotential for alkali plasmas taking into account the ion structure was made in the works \cite{bib48}, \cite{bib49}, where the Hellmann type pseudopotentials were used. In this work we use Hellmann-Gurskii-Krasko pseudopotential  model for electron-ion interactions and its modified version for ion-ion interactions. There is also a high interest to construct a pseudopotential model of particle interaction in dense plasmas taking into account not only the quantum-mechanical effects including the ion shell structure at short distances but also the screening field effects. In the work \cite{bib73} the screened Hellmann-Gurskii-Krasko potential was derived using Bogolyubov's method as described e.g. in \cite{bib59}, \cite{bib31} and \cite{bib53}.\\ 
\indent We consider Li$^+$, Na$^+$, K$^+$, Rb$^+$, Cs$^+$ and Be$^{2+}$ two-component (TCP) plasmas with the charges $z e_-=-e_+$ and masses $m_i>>m_e$ and the densities $n_e=z n_i$ ($z=1, 2$). We will calculate here the TCP static and dynamic structure factors, including quantum effects and the ion shell structure using the Hellmann-Gurskii-Krasko pseudopotential (HGK). To determine the static and dynamic structure factors we use the screened Hellmann-Gurskii-Krasko potential obtained in \cite{bib73}. The method which is used for the calculation of the static structure factor is the TCP hypernetted-chain approximation developed for the case of absence of the local thermodynamic equilibrium (LTE) by Seuferling et al \cite{bibSeuf} and further discussed and extended for SSF by Gregori et al. \cite{bibGreg2006} while for the dynamic - the method of moments applied to two-component plasmas \cite{bibAdam83}, \cite{bibAdam93}. We would like to underline again that the inclusion of both components into the theory and a correct account of the short-range electron-ion interactions, is essential for the understanding of the structure factors in plasmas. \\

\section{Pseudopotentials taking into account the ion structure. Hellmann-Gurskii-Krasko potential}
\label{sec:1}
\begin{figure}
\resizebox{1\columnwidth}{!}{%
  \includegraphics{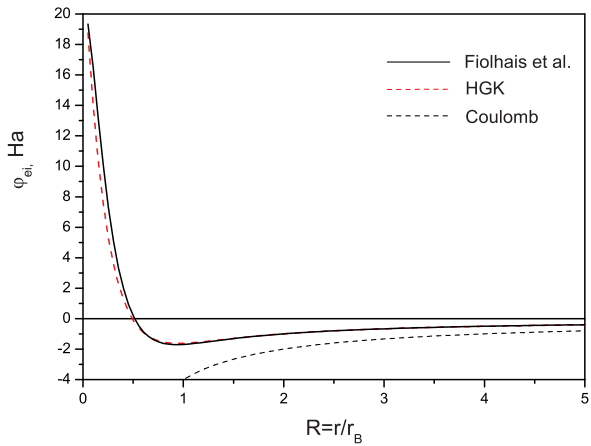}
}
\caption{Comparisons among the $e-i$ HGK, Fiolhais et al. and Coulomb potentials for Be$^{2+}$ (in atomic units).}
\label{Fig:21}
\end{figure}

\begin{figure*}[htb]
\resizebox{1\columnwidth}{!}{%
\includegraphics{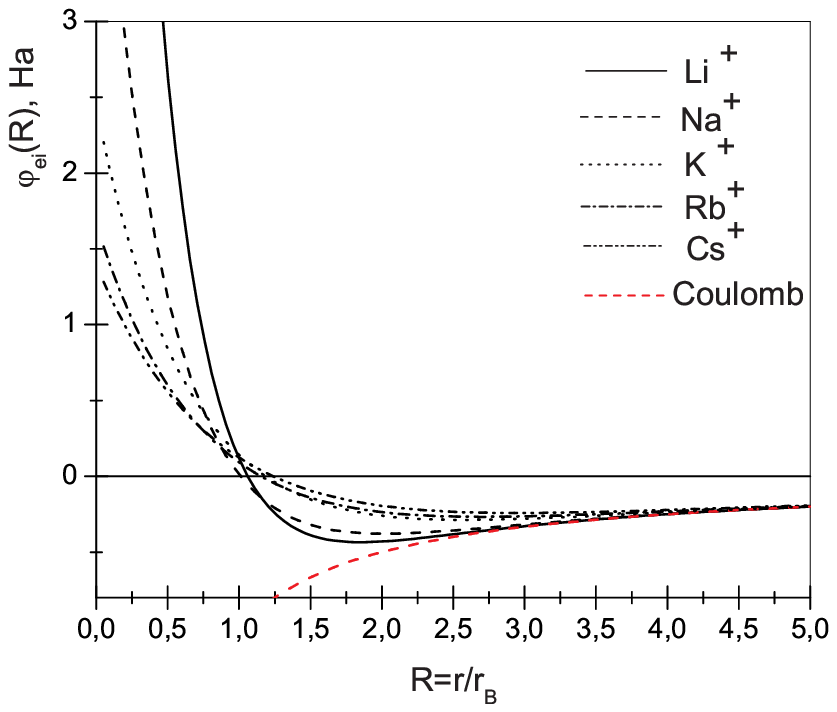}~a)}
\hfil
\resizebox{1\columnwidth}{!}{%
\includegraphics{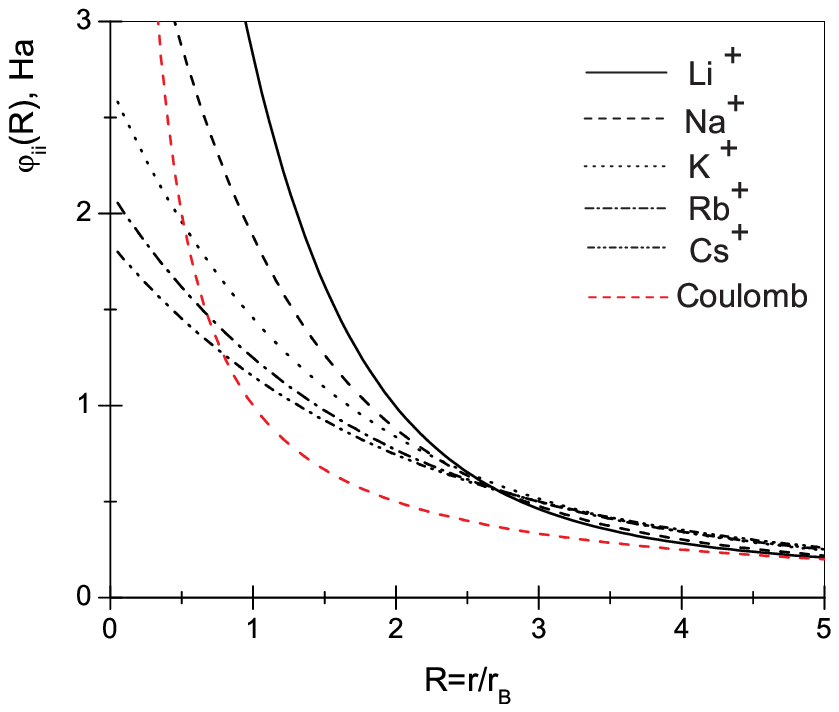}~b)}
\caption{ Comparisons among the HGK potentials of different alkali plasmas (in atomic units).}
\label{Fig:1}
\end{figure*}

\indent Clearly, the simple Coulomb law is not applicable to describe the forces between the charges in alkali plasmas since, at small distances, there are strong deviations from Coulomb\textquoteright s law due to the influence of core electrons. \\
\indent In many problems of atomic and molecular physics one can divide the electrons of the system into valence and core electrons. A number of important physical properties are determined by the valence electrons. In a series of pioneering papers Hellmann attempted to develop a model in which the treatment of atoms and molecules in alkali plasmas is reduced to the treatment of valence electrons \cite{bib72}. Hellmann demonstrated that the Pauli exclusion principle for the valence electrons can be replaced by a nonclassical potential (\emph{repulsive potential}) which is now called the \emph{pseudopotential}. Hellmann's idea was to replace the requirement of orthogonality of valence orbital to the core orbitals by the pseudopotential and it was this idea which made the respective mathematical calculations much simpler. \\
\indent For the actual purpose of atomic and molecular calculations Hellmann suggested a simple analytic formula. Let $\varphi$ be the sum of electrostatic, exchange, and polarization potentials, representing the interaction between a valence electron and the core of an atom. Let $\varphi_p$ be the \emph{repulsive potential}. The \emph{Hellmann} potential $\varphi_H=\varphi+\varphi_p$ may be expressed as:
 \begin{equation}\label{100}
{\varphi^H}_{ei}(r) = -\frac{z e^2}{4\pi\varepsilon_0 r}+\frac{e^2}{4\pi\varepsilon_0 r}A \exp(-\alpha r),
\end{equation}
Here $z$ is the ionic charge of the core; that is, if the nucleus contains $Z$ positive charges and the core contains $N$ electrons then $z=Z-N$. The constants $A$ and $\alpha$ are determined from the requirement that the potential $\varphi_H$ should reproduce the energy spectrum of the valence electron as accurately as possible. Later several modifications were introduced by Schwarz, Bardsley etc. into the determination of the Hellmann potential parameters without changing the basic analytic form of the potential \cite{bib101}. For instance, Schwarz improved the determination of the potential parameters of  Be$^+$, Mg$^+$, Ca$^+$, Sr$^+$, Zn$^+$ and Li, Na, K, Rb, Cu periodic families consistently obtaining a better fit to the empirical energy levels \cite{bib600}.\\
\indent However, all above pseudopotentials have one drawback. They are usually described in the $\vec{r}$ space by a discontinuous function or have a relatively hard core as in the case of Hellmann's potential. As a result, their Fourier transforms (formfactors), $\varphi(k)$, at $k\to \infty$, do not guarantee the convergence of series and integrals of the perturbation theory. Alternatively, Gurskii and Krasko constructed a pseudopotential model excluding this shortcoming by introducing a continuous in the $\vec{r}$ space pseudopotential. To include the smoothness of the obtained electron density distribution in an atom, Gurskii and Krasko proposed the following electron-ion model pseudopotential \cite{bib51}, \cite{bib54}:
 \begin{eqnarray}\label{28}
{\varphi^{HGK}}_{ei}(r) &=& -\frac{z e^2}{4\pi\varepsilon_0 r}\left[1-\exp\left(-\frac{r}{{R_C}_{ei}}\right)\right]\nonumber\\&&+\frac{z e^2}{4\pi\varepsilon_0}\frac{a}{{R_C}_{ei}}\exp\left(-\frac{r}{{R_C}_{ei}}\right),
\end{eqnarray}
where ${R_C}_{ei}={r_C}_{ei} r_B$ and $a$ are determined experimentally using the ionization potential and the formfactor of the screened pseudopotential at the first nodes of the reciprocal lattice. The parameter ${r_C}_{ei}$ is defined as a certain radius characterizing the size of the region of internal electron shells. If such measurements are not available, the second condition is replaced by the constraint that the pressure $P=0$ at zero temperature in the equillibrium lattice. The magnitudes are given in the SI system of units. In this work values of $a$, ${r_C}_{ei}$ are taken from \cite{bib62}. Since the first two terms in (\ref{28}) are identical to the potential of Hellmann \cite{bib72}, we call this pseudopotential the Hellmann-Gurskii-Krasko potential. The results of calculation of both the bound energy and the phonon spectra using the Hellmann-Gurskii-Krasko (HGK) potential were found in a good agreement with the experimental data and can be used in a wide range of investigation of thermodynamic properties of alkali plasmas. Unfortunately there are no available HGK parameters for the Be$^{2+}$ ion. That is why we looked for alternative $e - i$ potentials with the determined for Be$^{2+}$ parameters. It is the Fiolhais et al. pseudopotential \cite{bibFiol}: 
\begin{equation}\label{Fiolh}
{\varphi^{F}}_{ei}(r) = -\frac{z e^2}{4\pi\varepsilon_0 r_c}\frac{1}{R}\{1-(1+\beta R) \exp(-\alpha R)\}
\end{equation}
that we used to describe Be$^{2+}$ plasmas. Here $R=r/r_c$, $r_c$ being a core decay length, $\alpha> 0$, $\beta=(\alpha^3-2\alpha)/4(\alpha^2-1)$ and $A=\alpha^2/2-\alpha\beta$. In \cite{bibFiol} there are two possible choices of parameters: the ``universal'' and the ``individual''. We made a fit of the``universal'' parameters of HGK to the Fiolhais et al. pseudopotential, which are $a=3.72$, $r=0.22$. In \cite{bibFiol} the universal parameters were chosen to obtain the best agreement between calculated and measured structure factors of alkali metals. In Fig. \ref{Fig:21} the comparison between the electron-ion Fiolhais et al., HGK and Coulomb potentials for Be$^{2+}$ plasma are shown. One can easily see that HGK almost coincided with the Fiolhais et al. potential. In Fig. \ref{Fig:1}a we display the pseudopotentials $\varphi_{ei}(r)$ for different alkali plasmas and the HGK pseudopotentials $\varphi_{ii}(r)$ are represented in Fig. \ref{Fig:1}b; notice that $\varphi_{ei}(r)$ possess a minimum. The Hellmann type pseudopotentials for alkali plasmas were proposed in \cite{bib48}, \cite{bib49} and the Hellmann-Gurskii-Krasko model for the ion-ion interaction shown in Fig. \ref{Fig:1}b for alkali plasmas is the following:
\begin{eqnarray}\label{30}
{\varphi^{HGK}}_{ii}(r) &=& \frac{z^2 e^2}{4\pi\varepsilon_0 r}\left[1-\exp\left(-\frac{r}{{R_C}_{ii}}\right)\right]\nonumber\\&&+\frac{z^2 e^2}{4\pi\varepsilon_0}\frac{a}{{R_C}_{ii}}\exp\left(-\frac{r}{{R_C}_{ii}}\right),
\end{eqnarray}
The values of ${r_C}_{ii}$, $a$ are not given in literature, particularly, ${r_C}_{ii}$ is taken hypothetically as the doubled value of that taken for $e-i$ interaction ${r_C}_{ii}=2 {r_C}_{ei}$ taking in this way both ions cores (closed shells) into account. We will study this in more detail and compare with the hard-core potential described in \cite{bib49}. In Table \ref{tab:5} the parameters of the Hellmann-Gurskii-Krasko potential for alkali elements and the elements of the second periodic family are presented. We note that ${\varphi^{HGK}}_{ei}(r)$ potential describes the interaction of a valence electron with the corresponding ion core of a charge $z$ and radius ${R_C}_{ei}$, while ${\varphi^{HGK}}_{ii}(r)$ describes the interaction between two ion cores of a charge $z$ with the same radius ${r_C}_{ii}$. \\

\begin{figure*}[htb]
\resizebox{1\columnwidth}{!}{%
\includegraphics{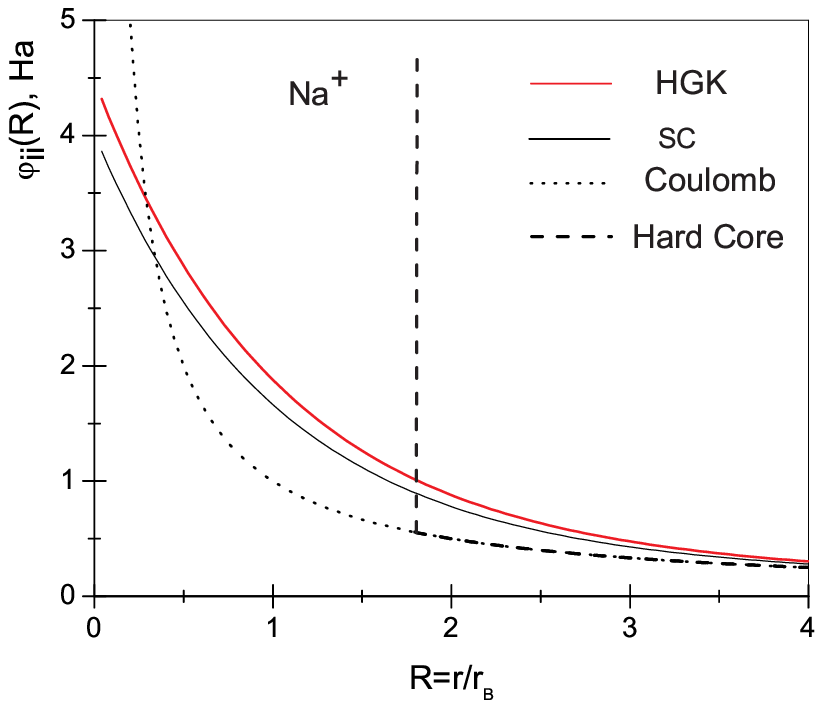}~a)}
\hfil
\resizebox{1\columnwidth}{!}{%
\includegraphics{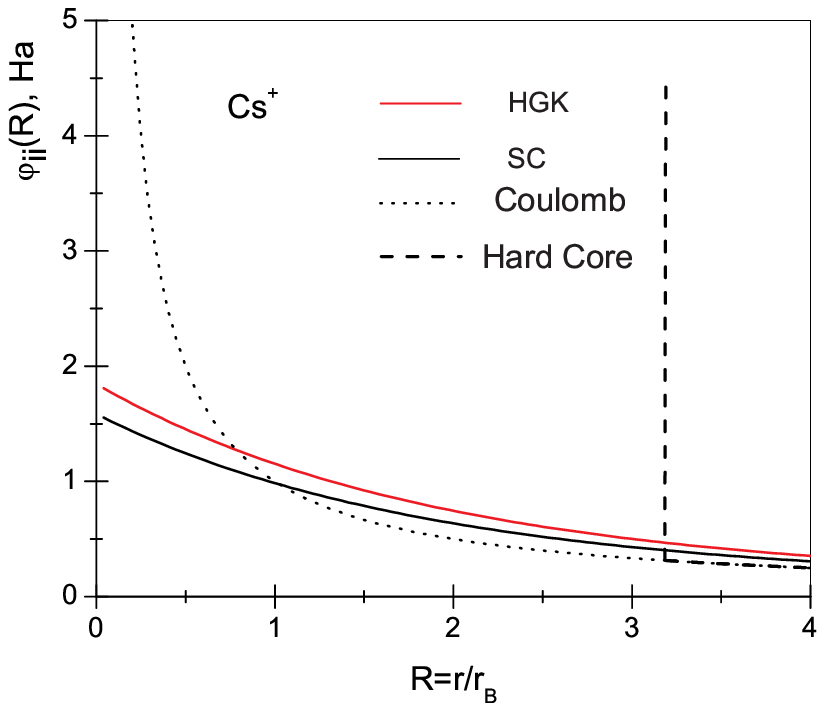}~b)}
\caption{ Comparison among the $i-i$ HGK, hard core, soft core and Coulomb potentials for (a) Na$^+$ and (b) Cs$^+$ plasmas (in atomic units).}
\label{Fig:22}
\end{figure*}

\begin{table*}
\caption{The parameters of the Hellmann-Gurskii-Krasko potential in atomic units. Here, in the case $z=2$ the given parameters correspond to the interaction of a double charged ion with an electron.}
\centering
\label{tab:5}\renewcommand{\arraystretch}{1.5}
\begin{tabular}{lllllllllll}\hline
\noalign{\smallskip}
&Li&Na&K&Rb&Cs&Be&Mg&Ca&Sr&Br\\ 
\noalign{\smallskip}
\hline\noalign{\smallskip}
$z$&1&1&1&1&1&2&2&2&2&2\\ 
a&5.954&3.362& 2.671&2.293& 2.214& 3.72& 2.588& 2.745& 2.575& 2.870 \\ 
$rc_{ei}$& 0.365 &0.487&0.689& 0.779& 0.871& 0.22& 0.427& 0.571& 0.644& 0.698 \\ 
$rc_{ii}$& 0.73 &0.974& 1.948 &1.558& 1.742 &0.44& 0.854& 1.142& 1.288& 1.396 \\ 
\noalign{\smallskip}\hline
\end{tabular}
\end{table*}
 In principle, the choice of the parameters for the ion-ion interaction should be based on methods similar to those in \cite{bib62}. Furthermore, our calculations led us to the conclusion that the potential is not sensitive to the $a$ parameter of the ion-ion interaction. That is why $a$ was taken the same as in the electron-ion potential. In Figures \ref{Fig:22}a, b the comparison between the HGK, hard core (HC) \cite{bib49} and soft core (SC) (eq. (5) in \cite{bibSad22009}) models is presented. \\
 \indent The pseudopotentials which are used in our calculations were originally developed for applications in the electronic band structure and binding energies in alkali metals. However the derivation used by Hellmann and his followers is basically working with wave functions of a few particles and not the multiparticle wave functions of the solid state. For this reason we cannot see strong arguments against  the two-particle interaction we employ in the plasma state. Of course this is a working assumption which needs further justification. Anyhow we are convinced that the application of pseudopotentials of Hellmann-type is much nearer to reality than the use of pure Coulomb potentials or hard-core potentials. Further, we would like to argue that the experimental investigations of alkali metals near the critical point did not evidence the existence of deep differences between the two particle interactions in the liquid and the gaseous state \cite{bib74}, \cite{bib75}. What is clearly different are the multi-particle interactions, however multi-particle effects are less relevant at the densities we consider here.

\subsection{Screening of the Hellmann-Gurskii-Krasko potential}
\label{sec:2}
\begin{figure*}
\resizebox{1\columnwidth}{!}{%
\includegraphics{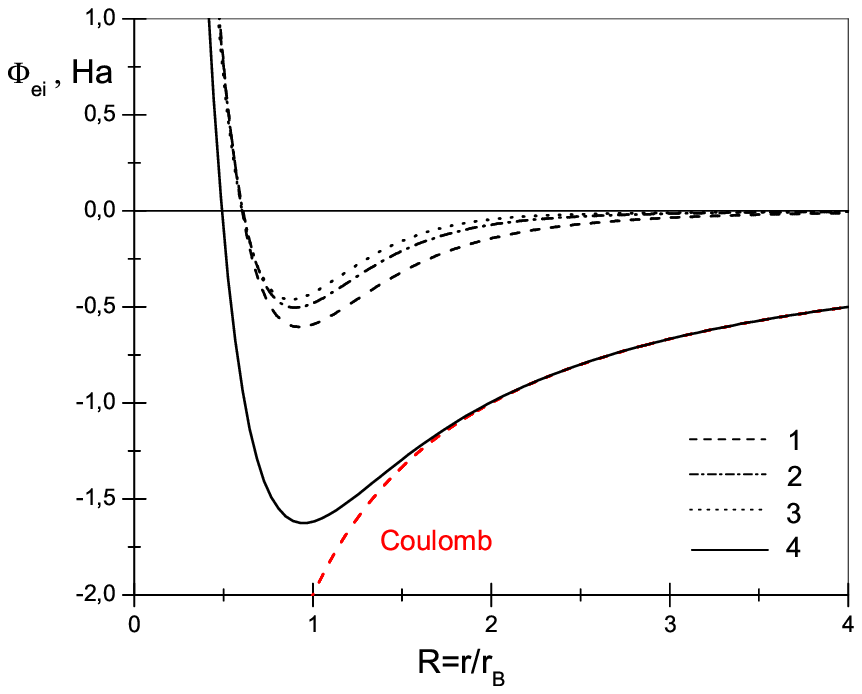}~a)}
\hfil
\resizebox{1\columnwidth}{!}{%
\includegraphics{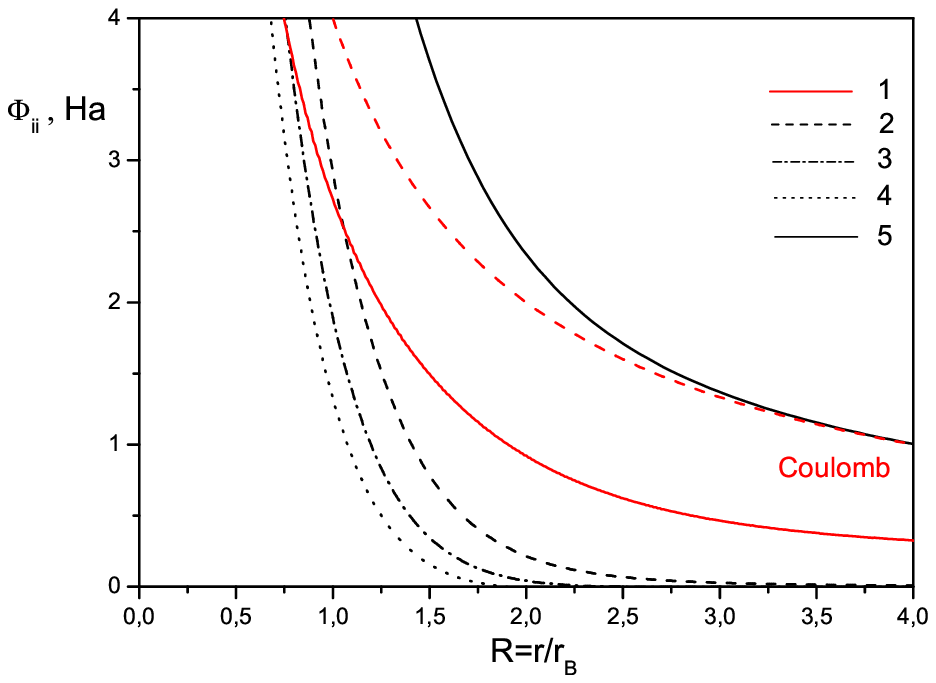}~b)}
\caption{ The screened e-i HGK $\Phi_{ei}$ and i-i HGK $\Phi_{ii}$ for beryllium plasma ($Be^{2+}$) (in atomic units). a)1: Screened HGK at $T_e=T_i=40 eV $,$T_e'=42.17 eV$, $\Gamma_{ee}=0.346$; 2: Screened HGK at $T_e=T_i=20eV $, $T_e'=24.06 eV$, $\Gamma_{ee}=0.606$, 3: Screened HGK at $T_e=T_i=13eV $, $T_e'=18.65$, $\Gamma_{ee}=0.782$, 4: HGK; b) 1: S.S. Dalgic et al. at $T_e=T_i=13eV $, $T_e'=18.65$, $\Gamma_{ee}=0.782$,2: Screened HGK at $T_e=T_i=40 eV $, $T_e'=42.17 eV$, $\Gamma_{ee}=0.346$; 3: Screened HGK at $T_e=T_i=20eV $, $T_e'=24.06 eV$, $\Gamma_{ee}=0.606$,4: Screened HGK at $T_e=T_i=13eV $, $T_e'=18.65$, $\Gamma_{ee}=0.782$, 5: HGK. The plasma parameters are described in the section \ref{sec:3}}
\label{Fig:HGKeiG165G073}
\end{figure*}

\indent Effective potentials simulating quantum effects of diffraction and symmetry \cite{{bib36}} - \cite{bib58} as well as significantly improved potentials \cite{bib37}-\cite{bib29} are widely used to determine the thermodynamic and transport properties of semiclassical fully ionized plasmas. In particular, Deutsch and co-workers \cite{bib57}, \cite{bib58} obtained the following effective interaction potential of charged particles in a plasma medium:
 \begin{equation}\label{36}
 \varphi_{ee}=\frac{e^2}{4\pi\varepsilon_0 r} \left[1-\exp\left(-\frac{r}{\lambda_{ee}}\right)\right]+k_B T \ln2 \exp\left(-\frac{r^2}{\lambda_{ee}^2 \ln2}\right),
  \end{equation}
where $\lambda_{ee}=\hbar/\sqrt{m_{e}k_BT}$ is the electron thermal de-Broglie wavelength. In the work \cite{bib53} it was proposed to use the $e-e$ effective potential (\ref{36}), along with the corresponding $e-i$ and $i-i$ potentials at short distances and the screened potential, treating three particle correlations, at large ones. The transition from one potential curve to another was realized at the intersection point by the spline-approximation method. \\
\indent The pseudopotential model (\ref{36}) was developed only for highly ionized plasmas. Since most experimental data available refer to partially ionized plasmas at moderate temperatures when the ions partially retain their inner shell, it is of a high interest to construct the pseudopotential model which takes into account not only the quantum-mechanical and screening field effects but also the ion shell structure. In order to include the screening effects, in \cite{bib73} we applied the method developed in \cite{bib31} and \cite{bib53}. In \cite{bib53} the authors suggested the classical approach based on the chain of Bogolyubov equations \cite{bib59} for the equilibrium distribution functions where the potential (\ref{36}) was taken as a micropotential. In \cite{bib73} the Fourier transforms of the screened HGK were derived,  eq. (11)-(14) therein, using the $e-i$, $i-i$ Hellmann-Gurskii-Krasko pseudopotentials (\ref{28}), (\ref{30}) and $e-e$ Deutsch potential (\ref{36}) as the micropotentials. \\
\indent The screened HGK potential was obtained in \cite{bib73} in the following way. In the Fourier space the system of integro-differential equations which determines the screened HGK potential turns into the system of linear algebraic equations:
\begin{equation}\label{39}
\Phi_{ab}(k)=\varphi_{ab}(k)-\frac{1}{k_BT}\left[n_e\varphi_{ae}(k)\Phi_{eb}(k)+n_i\varphi_{ai}(k)\Phi_{ib}(k)\right]
\end{equation}
where $a,b=i,e$. Solving the system (\ref{39}) for two-component plasma one can derive the following expressions for the Fourier transform $\Phi_{ab}(k)$\footnotemark of the pseudopotential $\Phi_{ab}(r)$:
\footnotetext[1]{A correction is made with respect to a misprint in \cite{bib73}: in eqs. (\ref{43}-\ref{45}) there should be $\sim\exp\left(-\frac{k^2}{4b}\right)$ instead of $\sim\exp\left(-\frac{k^2}{4b^2}\right)$}
\begin{equation}\label{42}
\Phi_{ei}(k)=\frac{z e^2}{\varepsilon_0 \Delta}\frac{(2a-1) {{R_c}_{ei}}^2 k^2-1}{k^2(1+k^2 {{R_c}_{ei}}^2)^2 },
\end{equation}
\begin{eqnarray}
&\Phi_{ee}(k)=\frac{e^2}{\varepsilon_0 \Delta}\left\{\frac{1}{k^2(1+k^2 {{\lambda}_{ee}}^2)}  \right.\nonumber \\  &+\frac{1}{k^4 {r_{Di}}^2}\left[\frac{(2a+1) {{R_c}_{ii}}^2 k^2+1}{(1+k^2 {{\lambda}_{ee}}^2)(1+k^2 {{R_c}_{ii}}^2)^2} - \left(\frac{(2a-1) {{R_c}_{ei}}^2 k^2-1}{(1+k^2 {{R_c}_{ei}}^2)^2 }\right)^2\right]\nonumber \\   &+\left. A\left( 1+\frac{(2a+1) {{R_c}_{ii}}^2 k^2+1}{k^2 {r_{Di}}^2(1+k^2 {{R_c}_{ii}}^2)^2}\right)\exp\left(-\frac{k^2}{4b}\right)\right\},
\label{43}
\end{eqnarray}
\begin{eqnarray}
&\Phi_{ii}(k)=\frac{ z^2 e^2}{\varepsilon_0 \Delta}\left\{ \frac{(2a+1) {{R_c}_{ii}}^2 k^2+1}{k^2 (1+k^2 {{R_c}_{ii}}^2)^2}\right.\nonumber \\  &+ \frac{1}{k^4 {r_{De}}^2}\left[\frac{(2a+1) {{R_c}_{ii}}^2 k^2+1}{(1+k^2 {{\lambda}_{ee}}^2)(1+k^2 {{R_c}_{ii}}^2)^2} - \left(\frac{(2a-1) {{R_c}_{ei}}^2 k^2-1}{(1+k^2 {{R_c}_{ei}}^2)^2 }\right)^2\right]\nonumber \\  & +\left. A\frac{(2a+1) {{R_c}_{ii}}^2 k^2+1}{k^2 {r_{De}}^2(1+k^2 {{R_c}_{ii}}^2)^2} \exp\left(-\frac{k^2}{4b}\right)\right\}.
\label{44}
\end{eqnarray}
\noindent Here $r_{De}$, $r_{Di}$ are the Debye radius of electrons and ions respectively, with $1/{r_{Di}}^2=z^2 e^2 n_i/(\varepsilon_0 k_BT)$,\\ \noindent $1/{r_{De}}^2= e^2 n_e/(\varepsilon_0 k_BT)$, $b=({{\lambda}_{ee}}^2 \ln2)^{-1}$,\\ \noindent $A=k_B T \ln2 \pi^{3/2} b^{-3/2}\varepsilon_0/e^2$ and 
\begin{eqnarray}
&\Delta=1+\frac{1}{k^2{r_{De}}^2(1+k^2 {{\lambda}_{ee}}^2)} + \frac{(2a+1) {{R_c}_{ii}}^2 k^2+1}{k^2 {r_{Di}}^2(1+k^2 {{R_c}_{ii}}^2)^2}  \nonumber \\& + \frac{1}{k^4{r_{De}}^2{r_{Di}}^2}\left[\frac{(2a+1) {{R_c}_{ii}}^2 k^2+1}{(1+k^2 {{\lambda}_{ee}}^2)(1+k^2 {{R_c}_{ii}}^2)^2} - \left(\frac{(2a-1) {{R_c}_{ei}}^2 k^2-1}{(1+k^2 {{R_c}_{ei}}^2)^2 }\right)^2\right]\nonumber \\  & +\frac{A}{{r_{De}}^2}\left( 1+\frac{(2a+1) {{R_c}_{ii}}^2 k^2+1}{k^2 {r_{Di}}^2(1+k^2 {{R_c}_{ii}}^2)^2}\right)\exp\left(-\frac{k^2}{4b}\right).
\label{45}
\end{eqnarray}
\indent The pseudopotential $\Phi_{ab}(r)$ can be restored from (\ref{42}-\ref{45}) by the Fourier transformation
\begin{equation}\label{46}
\Phi_{ab}(r)=\frac{1}{2\pi^2r}\int \Phi_{ab}(k) k \sin(kr)dk
\end{equation}
 The present approximation is restricted to the constraint $\Gamma_{ii} \lesssim 1$ due to the employment of the linearization procedure in the derivation of the resolved system of integro-differential equations.\\
 \indent In order to compare with the alternative $i - i$ pseudopotential models we considered the Dalgic potential. S.S. Dalgic et al. determined the screened ion-ion potential \cite{bibDalgic} on the basis  of the second order pseudopotential perturbation theory using the Fiolhais et al. potential ${\varphi^{F}}_{ei}(r)$:
\begin{equation}\label{ScreenDalg}
\Phi_{ii}^{D}(k)=\frac{4\pi z^2 e^2}{4\pi\varepsilon_0 k^2} + \chi(k) |{\varphi^{F}}_{ei}(k)|^2,
\end{equation}
where ${\varphi^{F}}_{ei}(k)$ is the pseudopotential local form factor. Here, we use instead of the Fiolhais potential the HGK $\Phi_{ii}^{HGK}(k)$ potential with the fitted to those of Fiolhais et al. potential parameters. Notice that $\chi(k)$ is the response of the electron gas, where the Lindhard response of a non-interacting degenerated electron gas $\chi^0(k)$ and the local field correction (LFC) $G(k)$ enter the expression in the standard way; the LFC accounts for the interactions between the electrons. We used the LFC which satisfies the compressibility sum rule at finite temperatures obtained by Gregori et al. for the strong coupling regime \cite{bibGreg2007}. Similar calculations have also been carried out by E. Apfelbaum who used the potential described by Dalgic et al. to calculate the SSF of Cs and Rb in the realm of the liquid-plasma transition \cite{bibApfel}. The author of \cite{bibApfel} showed that calculated data were in agreement with the measured SSF.\\
 \indent In Fig. \ref{Fig:HGKeiG165G073} the $e-i$ and $i-i$ HGK, screened HGK and $i-i$ Dalgic et al. potentials are presented for comparison. One can easily see that with the growth of $\Gamma_{ee}$ (defined in the section \ref{sec:3}) the curves shift in the direction of its low absolute values. We presume that this occurs due to the increasing role of screening effects. 

\subsection{Static Structure Factors}
\label{sec:3}
\begin{figure*}
\sidecaption
\begin{minipage}[t]{176mm}
\begin{minipage}[t]{88mm}
\resizebox{1\columnwidth}{!}{\includegraphics{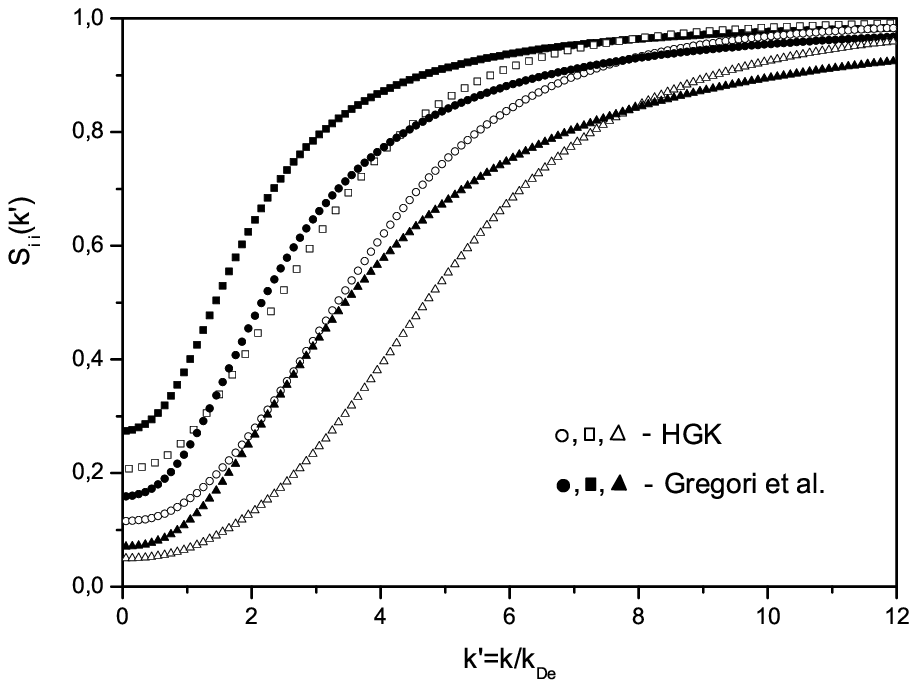}~a)}\\[3mm]%
\resizebox{1\columnwidth}{!}{\includegraphics{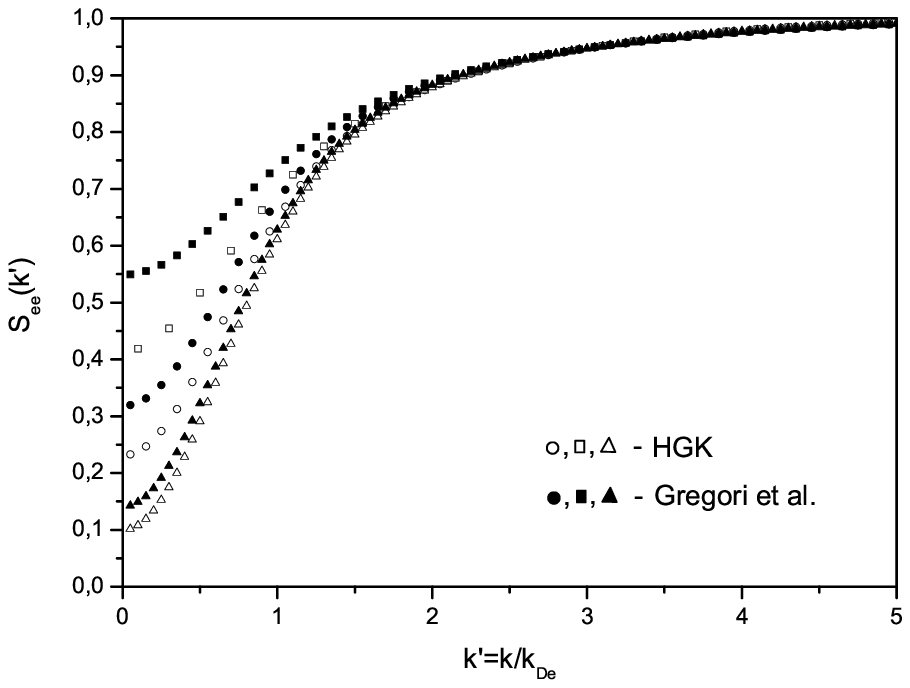}~b)}
\end{minipage}
\hfill%
\begin{minipage}[t]{88mm}
\resizebox{1\columnwidth}{!}{\includegraphics{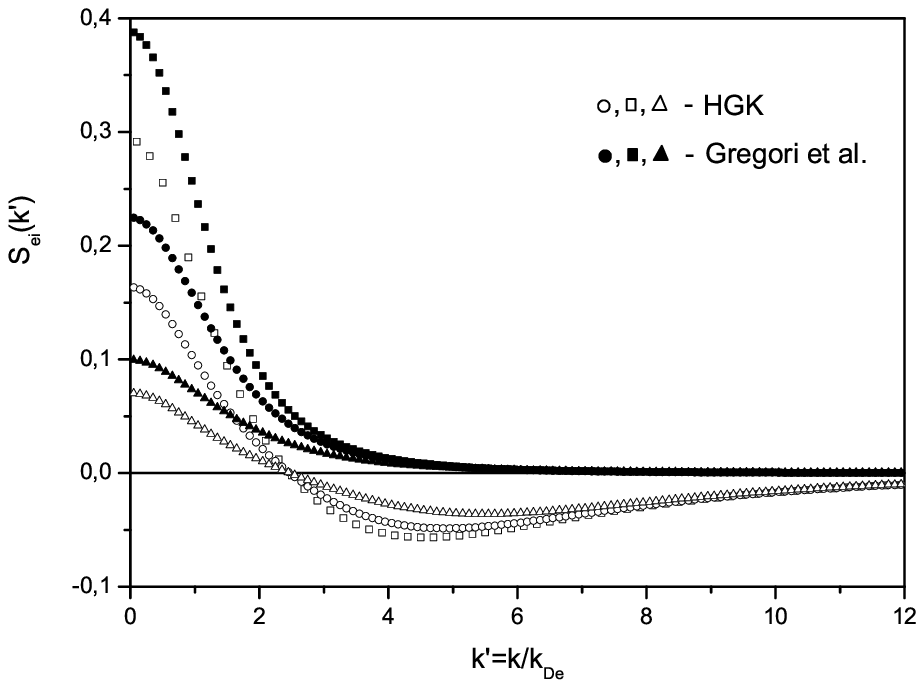}~c)}\\[3mm]%
\resizebox{1\columnwidth}{!}{\includegraphics{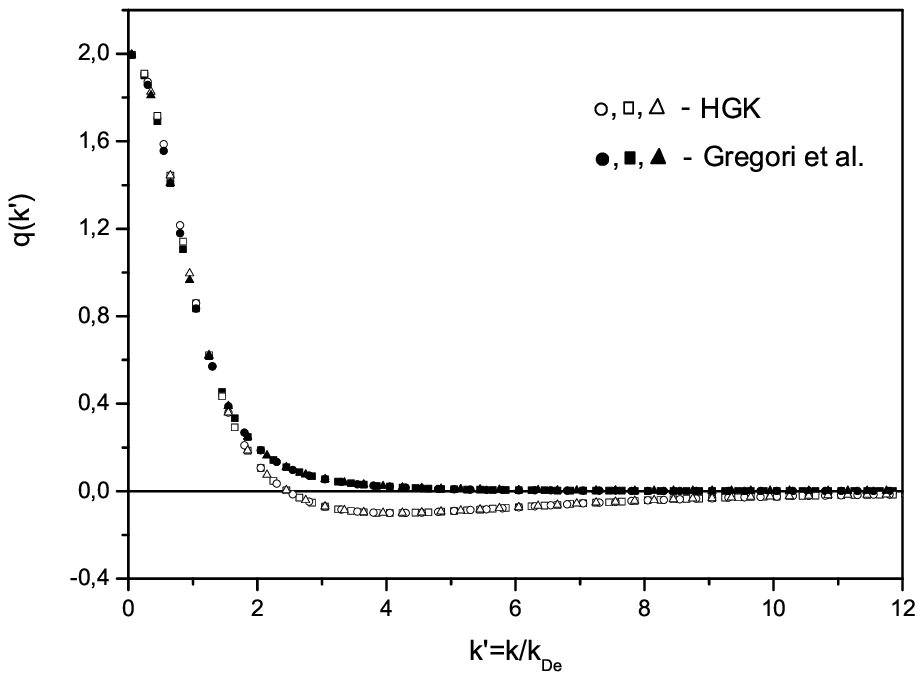}~d)}
\end{minipage}
\end{minipage}%
\caption{Static structure factors and screening charge $q(k')$ for $Be^{2+}$ plasma at $T_e=20eV$, $T_e'=24.06 eV$, $z=2$ and $n_e=2.5\times 10^{23} cm^{-3}$. The set of filled symbols  represents the screened Deutsch model obtained by Gregori et al.\cite{bibGreg2006} , while the set of hollow symbols - the screened HGK model. Squares: $T_i/T_e=1$ $(\Gamma_{ii}=2.31$, $\Gamma_{ee}=0.61)$. Circles: $T_i/T_e=0.5$ $(\Gamma_{ii}=4.63$, $\Gamma_{ee}=0.61)$. Triangles: $T_i/T_e=0.2$ $(\Gamma_{ii}=11.57$, $\Gamma_{ee}=0.61)$.}
\label{Fig:12}
\end{figure*}

\begin{figure}
\resizebox{1\columnwidth}{!}{
\includegraphics{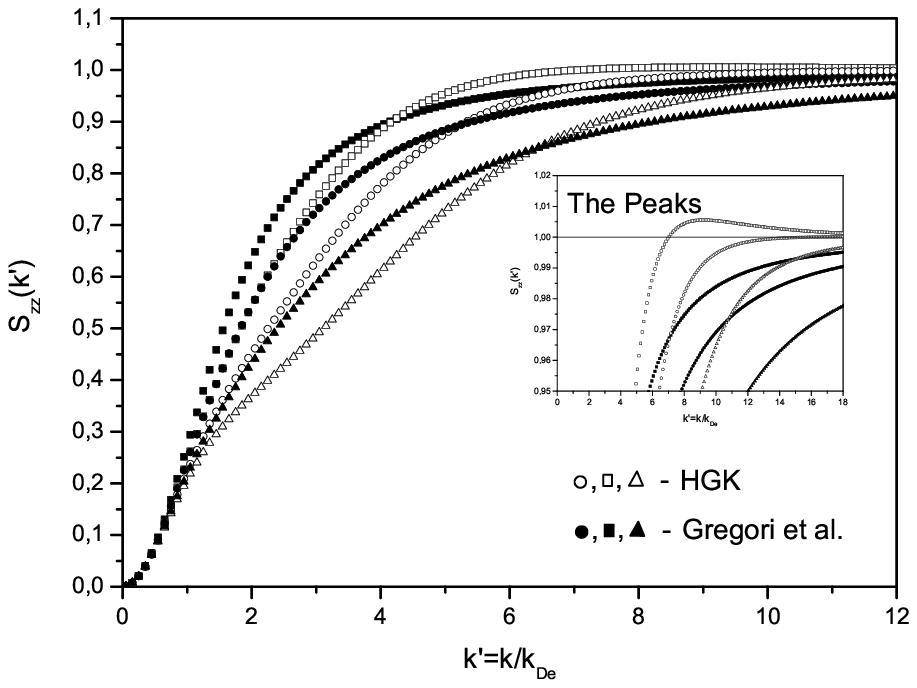}}
\caption{The charge-charge static structure factors $S_{zz}$ (\ref{Scc}) for a beryllium plasma with $n_e\approx 2.5 \cdot 10^{23} cm^{-3}$, $z\approx 2$, and $T_e=20 eV$, $T_e'=24.06 eV$. The set of filled symbols represents the screened Deutsch model obtained here on a base of Gregori et al.\cite{bibGreg2006} , while the set of hollow symbols - the screened HGK model. Squares: $T_i/T_e=1$ $(\Gamma_{ii}=2.31$, $\Gamma_{ee}=0.61)$. Circles: $T_i/T_e=0.5$ $(\Gamma_{ii}=4.63$, $\Gamma_{ee}=0.61)$. Triangles: $T_i/T_e=0.2$ $(\Gamma_{ii}=11.57$, $\Gamma_{ee}=0.61)$.}
\label{Fig:SccBe}
\end{figure}

\begin{figure*}
\sidecaption
\begin{minipage}[t]{176mm}
\begin{minipage}[t]{88mm}
\resizebox{1\columnwidth}{!}{\includegraphics{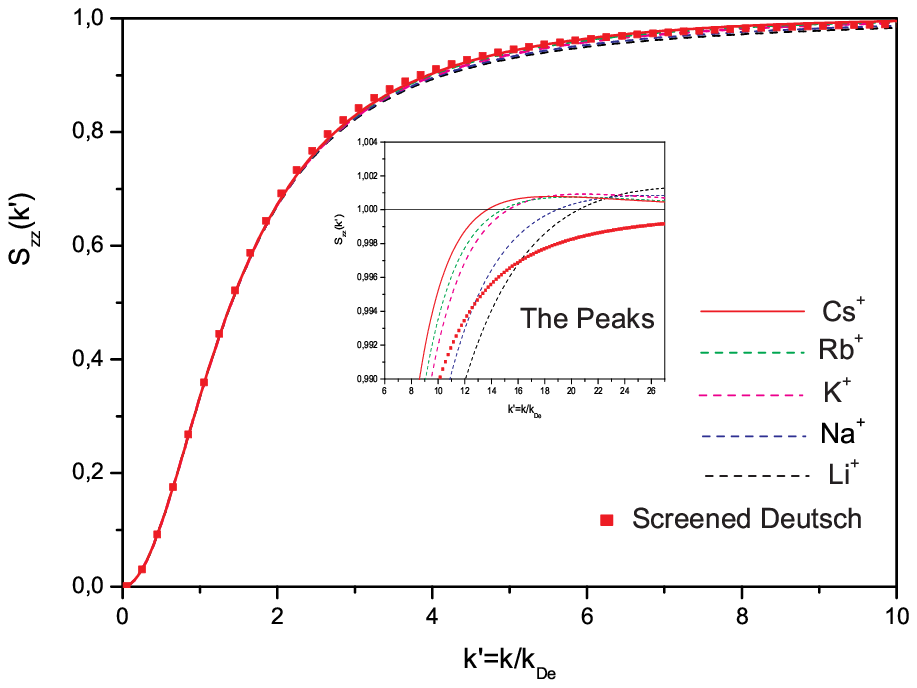}~a)}\\[3mm]%
\resizebox{1\columnwidth}{!}{\includegraphics{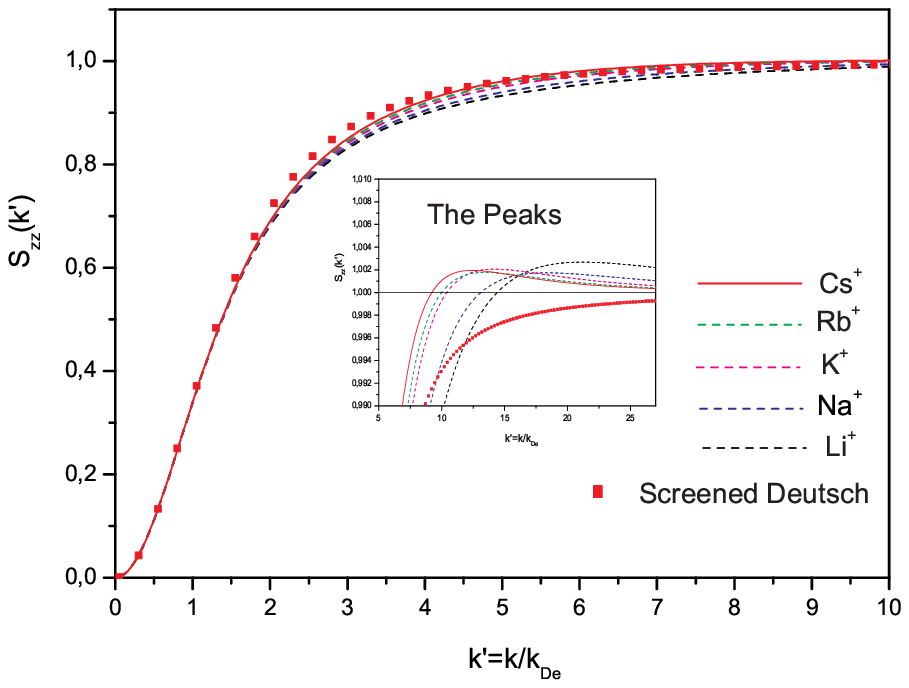}~b)}
\end{minipage}
\hfill%
\begin{minipage}[t]{88mm}
\resizebox{1\columnwidth}{!}{\includegraphics{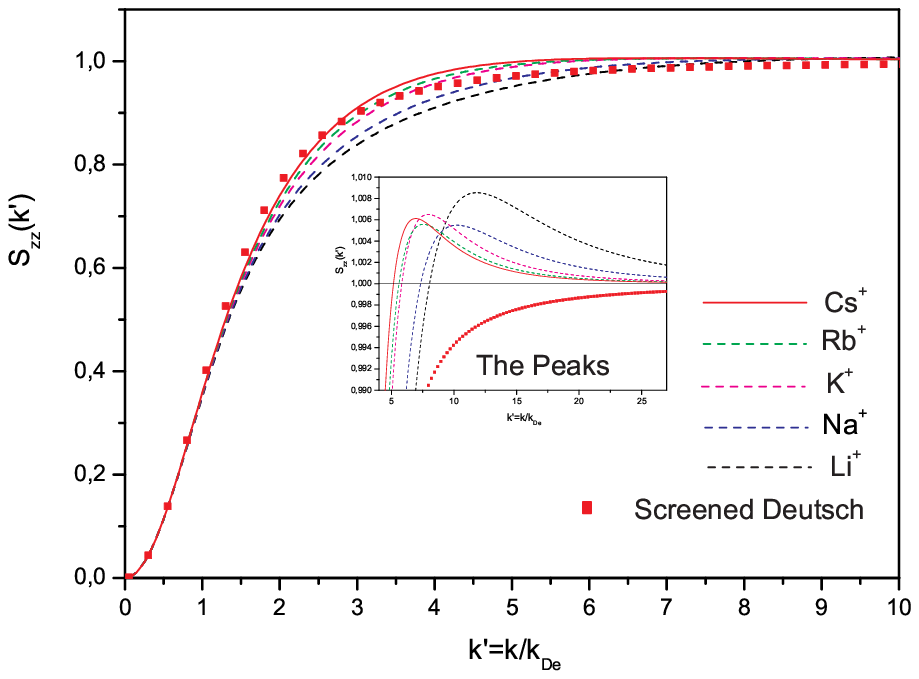}~c)}\\[3mm]%
\resizebox{1\columnwidth}{!}{\includegraphics{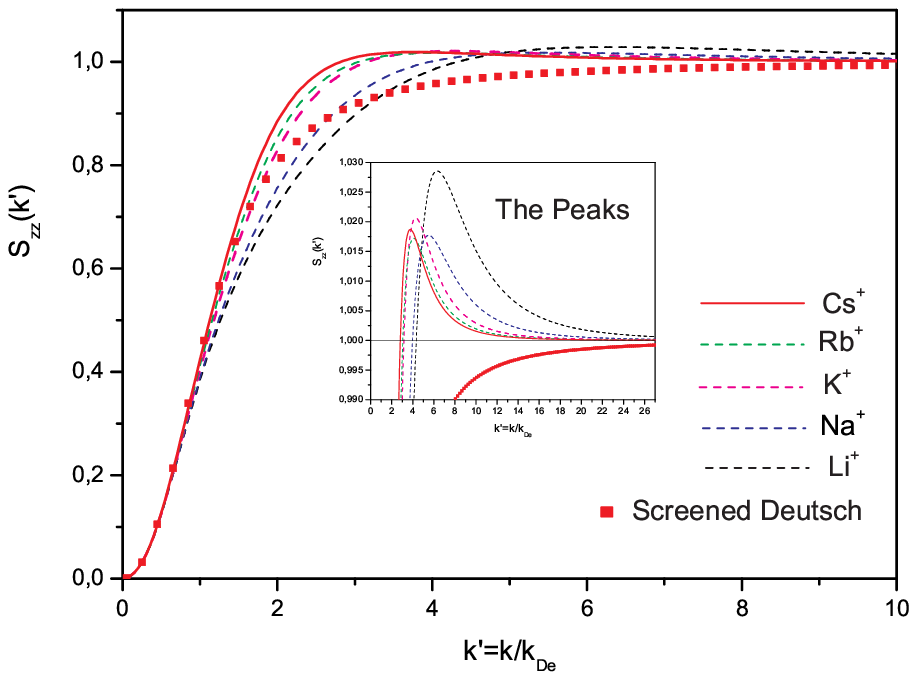}~d)}
\end{minipage}
\end{minipage}%
\caption{The charge-charge static structure factors $S_{zz}$ (\ref{Scc}) for alkali plasmas (Li$^+$, Na$^+$,K$^+$, Rb$^+$, Cs$^+$) in a frame of the screened HGK model and results obtained in the present work for hydrogen-like plasmas in a frame of the screened Deutsch model on a base of Gregori et al.\cite{bibGreg2006}. (a) $T_e=T_i=60000 K$, $T_e'=60204 K$, $\Gamma_{ee}=0.398$, $\Gamma_{ii}=0.399$; (b) $T_e=T_i=30000 K$, $T_e'=30407 K$, $\Gamma_{ee}=0.789$, $\Gamma_{ii}=0.8$; (c) $T_e=T_i=30000 K$, $T_e'=31471 K$, $\Gamma_{ee}=1.14$, $\Gamma_{ii}=1.2$; (d) $T_e=T_i=30000 K$, $T_e'=37806 K$, $\Gamma_{ee}=1.58$, $\Gamma_{ii}=2$. }
\label{Fig:SccG039G08G12G2}
\end{figure*}

Within the framework of the density response formalism for two-component plasmas, we calculated the screened HGK interaction potentials described above using the semiclassical approach suggested by Arkhipov et al. \cite{bib53}. Our approach is based on the HGK pseudopotential model for the interaction between the particles (charged spheres) to account for ion structure, quantum diffraction effects i.e., the Pauli exclusion principle and symmetry. Quantum diffraction is represented by the thermal de Broglie wavelength $\lambda_{rs}=\hbar/\sqrt{2 \mu_{rs}'k_BT_{rs}'}$ with $\mu_{rs}'=\frac{m_rm_s}{m_r+m_s}$ the reduced mass of the interacting pair $r-s$, and $r$ , $s=e$ (electrons) or $i$ (ions). The effective temperature $T_{rs}'$ is given by ,
\[T_{rs}'=\frac{m_rT_s'+m_sT_r'}{m_r+m_s},\]\\
where $T_e'=(T_e^2+T_q^2)^{1/2}$ with $T_q=T_F/(1.3251-0.1779\sqrt{r_s})$,  where $r_s=r_a/r_B$, $T_F=\hbar^2(3\pi^2n_e)^{2/3}/(2k_B m_e)$ and $T_i'=(T_i^2+\gamma_0 T_D^2)^{1/2}$, $T_D=\Omega_{pi}\hbar/k_B$,  $\gamma_0=0.152$ is the Bohm-Staver relation for the Debye temperature with $\Omega_{pi}^2=\omega_{pi}^2/(1+k_{De}/k^2)$, $\omega_{pi}=\sqrt{z e^2 n_e/(\varepsilon_0 m_i)}$ with $m_i$ being the ion mass, $k_ {De}=\sqrt{e^2 n_e/(\varepsilon_0 k_BT_e')}$ is the Debye wave number for the electron fluid ($T_D\approx 0.16eV$, $T_F\approx 14.5eV$ for $Be^{2+}$). This definition of the effective temperature allows to extend the fluctuation-dissipation theorem to nonequilibrium systems and is the input value for the partial static structure factors $S_{rs}(k)$, the Fourier transform of the pair distribution functions $h_{rs}(r)=g_{rs}(r)-1$ \cite{bibSeuf}:\\
\begin{equation}\label{Srsgrs}
S_{rs}(k) = \delta_{rs}+\sqrt{n_rn_s}\tilde{h}_{rs}(k)\simeq\delta_{rs}-\frac{\sqrt{n_rn_s}}{k_BT_{rs}'}\Phi_{rs}(k),
\end{equation}
\indent Since, in the Debye model, the phonon modes with wavelengths up to a fraction of the lattice spacing are considered, in \cite{bibGreg2006} it is set $k\equiv k_{max}=(2/z)^{1/3} k_F$ with $k_F=(3\pi^2 n_e)^{1/3}$ Fermi wave number. Due to the large mass difference between ions and electrons, $T_{ei}'=T_{ee}'$. All the parameters considered here are beyond the degeneration border $(n_e \lambda_{ee}^3<1)$. \\
\indent The partial static structure factors of the system are defined as the static (equal-time) correlation functions of the Fourier components of the microscopic partial charge densities \cite{bibHansen74}:\\
\begin{equation}\label{SrsRho}
S_{rs}(k ) = \frac{1}{N}<\rho^{r}(\vec{k})\rho^{s}(-\vec{k})>,
\end{equation}
where $N$ number of ions and
\begin{equation}\label{rho}
\rho^{r}(\vec{k}) =\sum_{i=1}^{N}\exp{(\imath \vec{k}\cdot {\vec{r}_i}^r )}.
\end{equation}
\noindent A linear combination of the partial structure factors which is of high importance, is the charge-charge structure factor defined as \\
 \begin{eqnarray}\label{SzzRho}
 S_{zz}(k)&=&\frac{1}{N_e+z N_i}<\rho^{z}(\vec{k})\rho^{z}(-\vec{k})> \nonumber \\
 &=& \frac{S_{ee}(k)-2\sqrt{z}S_{ei}(k)+z S_{ii}(k)}{2},
 \end{eqnarray}
 where $\rho^{z}=\rho^{i}(\vec{k})-\rho^{e}(\vec{k})$.\\
 \indent As described in \cite{bibGreg2006} by Gregori et al., the fluctuation-dissipation theorem may still be a valid approximation even under nonequilibrium conditions if the temperature relaxation is slow compared to the electron density fluctuation time scale. A common condition in experimental plasmas for this to occur is when $m_i>>m_e$ so that the coupling between the two-components takes place at sufficiently low frequencies. Using a two-component hypernetted chain (HNC) approximation scheme, Seuferling et al. \cite{bibSeuf} have shown that the static response under the conditions of the non-LTE  takes the form:\\
\begin{equation}\label{53}
S_{rs}(k ) = \delta_{rs}-\frac{\sqrt{n_rn_s}}{k_BT_{rs}'}\Phi_{rs}(k)-\delta_{er}\delta_{es}(\frac{T_{e}'}{T_{i}'}-1)\frac{|q(k)|^2}{z}S_{ii}(k) 
\end{equation}
where $q(k)=\sqrt{z}S_{ei}(k)/S_{ii}(k)$ and for $\Phi_{rs}$ the expression (\ref{42}-\ref{45}) was used.\\
\indent In Figures \ref{Fig:12} (a), (b), (c), (d) the static structure factors $S_{rs}(k)$ for a beryllium plasma with the introduced above different temperatures $T_i=T_e$, $T_i=0.5\cdot T_e$, $T_i=0.2\cdot T_e$ and the coupling parameters $\Gamma_{ee}=e^2/(4\varepsilon_0 k_BT_e'r_{ee})$, $\Gamma_{ii}=z^2e^2/(4\varepsilon_0 k_BT_i' r_{ii})$ with $r_{ii}=(3/4\pi n_i)^{1/3}$, $r_{ee}=(3/4\pi n_e)^{1/3}$ are shown. For typical conditions found in laser plasma experiments with solid density beryllium, we have $n_e\approx 2.5 \cdot 10^{23} cm^{-3}$ and $z\approx 2$. This gives $T_F\approx 14.5 eV$ and $T_D\approx 0.17 eV$. In Fig. \ref{Fig:12} (c) a minimum arises which defines the size of an ion core. Notice that in the Figure the minimum becomes less pronounced when the coupling increases. \\ 
 \indent In a screened OCP the effective response of the medium is described by the charge-charge correlation function  as given by Gregori et al. \cite{bibGreg2007}:\\
\begin{equation}\label{Scc}
S_{zz}^{G}(k)=\frac{S_{ee}(k)+z S_{ii}(k)-2 \sqrt{z} S_{ei}(k)}{1+z}
\end{equation}
\indent It is of high interest to study the influence of the ion structure on the static structure factors. For this reason in Figures \ref{Fig:12} (a)-(d) and further we compare the SSF, obtained from equations (\ref{53}) and (\ref{Scc}) with the help of the screened HGK potentials (\ref{42}-\ref{45}), with the corresponding SSF obtained using the screened Deutsch potential \cite{bib53}, on the basis of the TCP hypernetted-chain approximation developed for the case of non-LTE by Seuferling et al. \cite{bibSeuf} and further extended for DSF by Gregori et al. \cite{bibGreg2006}. \\
\indent In Fig. \ref{Fig:SccBe} the static charge-charge structure factor (\ref{Scc}) for a beryllium plasma with $n_e\approx 2.5 \cdot 10^{23} cm^{-3}$, $z\approx 2$, $T_e=20 eV$ and $T_i=T_e$, $T_i=0.5\cdot T_e$, $T_i=0.2\cdot T_e$ is shown.
 In Figures \ref{Fig:SccG039G08G12G2} (a) - (d) we compare our results on the charge-charge SSF (\ref{Scc}) for alkali plasmas within the \textbf{screened HGK model} with the results obtained in the present work for alkali (hydrogen-like point charges (\textbf{HLPC})) plasmas considered within the \textbf{screened Deutsch model} for various values of density and temperature. All curves obtained  within the screened Deutsch model converge due to the negligible influence of an alkali ion mass on the wavelength $\lambda_{ab}$ scale entering the equations in \cite{bib53}. As one can easily see with the growth of coupling the peaks become more pronounced and the difference among the curves becomes significant. We see that strong coupling and the onset of short-range order manifest themselves in $S_{zz}$ as a first localized peak, shown in an amplified scale, for different values of $k'$ for every alkali species, and the position of the peaks shifts in the direction of small values of $k'$. This phenomenon was also reported in \cite{bibGreg2007}.\\

\section{The dynamic structure factor: the moment approach}
\label{sec:4}

\begin{figure*}
\resizebox{1\columnwidth}{!}{%
\includegraphics{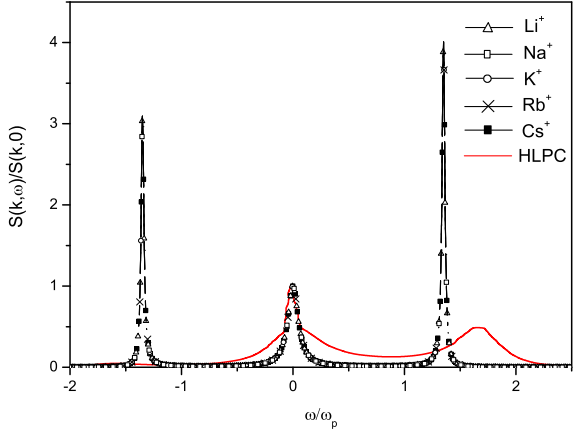}~a)}
\hfil
\resizebox{1\columnwidth}{!}{%
\includegraphics{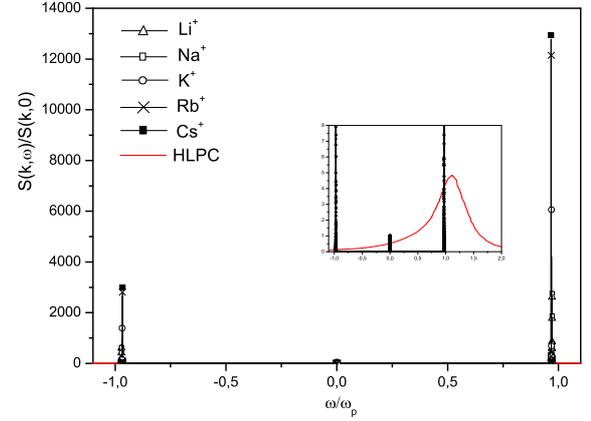}~b)}
\caption{Comparisons among the normalized dynamic structure factors of alkali plasmas (Li$^+$, Na$^+$, K$^+$, Rb$^+$, Cs$^+$) and the results of \cite{bibAdam93} for the HLPC model at $k=0.767/ r_{ee}$, (a) $T=30000 K$, $n_e=1.741 \cdot 10^{20} cm^{-3}$, $\Gamma_{ee}=0.5$ and (b) $T=30000 K$, $n_e=1.11 \cdot 10^{22} cm^{-3}$, $\Gamma_{ee}=2$. As the length scale we use the electron plasma frequency $\omega_{p}= n_e e^2/\varepsilon_0 m_e$.}
\label{Fig:206}
\end{figure*}

\begin{figure*}
\resizebox{1\columnwidth}{!}{%
\includegraphics{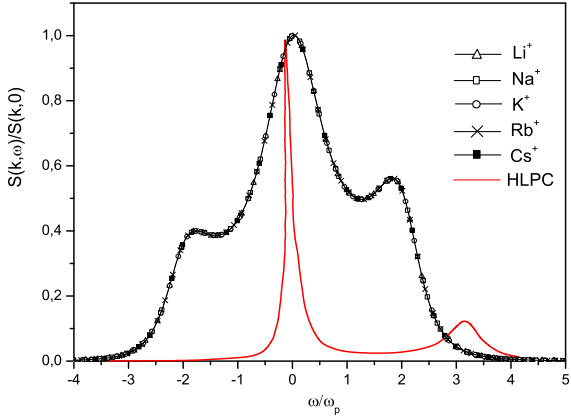}~a)}
\hfil
\resizebox{1\columnwidth}{!}{%
\includegraphics{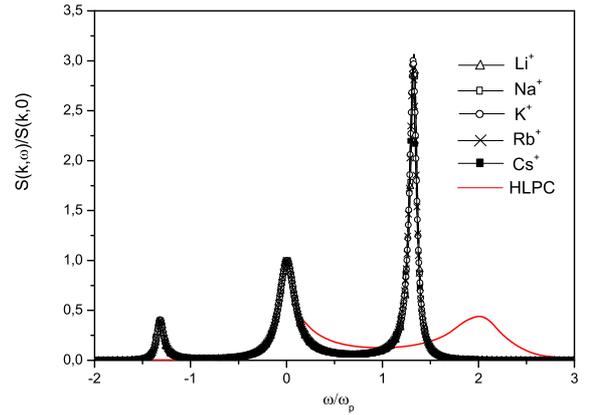}~b)}
\caption{Comparisons among the normalized dynamic structure factors of alkali plasmas (Li$^+$, Na$^+$, K$^+$, Rb$^+$, Cs$^+$) and the results of \cite{bibAdam93} for the HLPC model at $k=1.534/ r_{ee}$, (a) $T=30000 K$, $n_e=1.741 \cdot 10^{20}$, $\Gamma_{ee}=0.5$ and (b) $T=30000 K$, $n_e=1.11 \cdot 10^{22} cm^{-3}$, $\Gamma_{ee}=2$. As the length scale we use the electron plasma frequency $\omega_{p}= n_e e^2/\varepsilon_0 m_e$.}
\label{Fig:207}
\end{figure*}

Extensive molecular-dynamic simulations of Coulomb systems at a complete thermal equilibrium over a wide range of variation of the coupling parameter $\Gamma_{ii}$ ($z=1$) and $\Theta=E_F/k_BT $ ( $E_F$ is the Fermi energy) have been carried out by Hansen et al \cite{bibHansen74}. Hansen et al. studied dynamic and static properties of one-(OCP) and two-component plasmas and binary ionic mixtures. A new ``moment approach'' based on exact relations and sum rules was suggested in \cite{bibAdam83} in order to calculate dynamic characteristics of OCP and of the charge-charge dynamic structure factor of model semiquantal two-component plasmas. This approach proved to produce good agreement with the MD data of Hansen et al. Let 
\begin{equation}\label{rhot}
\rho^{r}(\vec{k},t) =\sum_{i=1}^{N}\exp{(\imath \vec{k}\cdot {\vec{r}_i}^r(t))}.
\end{equation}
be the Fourier components of the time dependent microscopic density of species $r$ \cite{bibEbelOrtner9897}. The corresponding dynamic structure factors are the Fourier transforms of the density-density time correlation functions given by 
\begin{equation}\label{SrsRhot}
S_{rs}(k, \omega ) = \frac{1}{2\pi N}<\rho^{r}(\vec{k},t)\rho^{s}(-\vec{k},0)>.
\end{equation}
 Alternatively, the charge-charge dynamic structure factor $S_{zz}(k,\omega)$ can be defined via the fluctuation-dissipation theorem (FDT) \cite{bibAdam93} as  
\begin{equation}\label{54}
S_{zz}(k,\omega) = -\frac{\hbar Im \varepsilon^{-1}(k,\omega)}{\pi\Phi(k)[1-\exp{(-\beta\hbar\omega)}]},
\end{equation}
where $\Phi(k)=e^2/\varepsilon_0 k^2$ and $\varepsilon^{-1}(k,\omega)$ is the inverse longitudinal dielectric function of the plasma. The
charge-charge dynamic structure factor is directly related to the charge-charge static structure factor as follows : \\
\begin{eqnarray}\label{Szz}
 S_{zz}(k)&=&\frac{1}{n_e+z n_i}\int_{-\infty}^{\infty}S_{zz}(k,\omega)d\omega \nonumber \\
 &=& \frac{S_{ee}(k)-2\sqrt{z}S_{ei}(k)+z S_{ii}(k)}{2},
 \end{eqnarray}
 where $T_{e}'=T_{i}'=T_e=T$, $T_{ei}'=T_{ee}'=T_{e}'$, $n_e=z n_i$ ($z=1$ for hydrogen-like plasmas). \\
\indent In order to construct the inverse longitudinal dielectric function within the moment approach one needs to consider the frequency moments of the loss function \\ 
$- Im \varepsilon^{-1}(k,\omega)/\omega$:
\begin{equation}\label{55}
C_{\nu}(k) = -\pi^{-1}\int_{-\infty }^{\infty}\omega^{\nu-1} Im \varepsilon^{-1}(k,\omega)d\omega,
\end{equation}
\indent Then the Nevanlinna formula of the classical theory of moments \cite{bibAkhieser} expresses the response function
\begin{equation}\label{56}
\varepsilon^{-1}(k,\omega)=1+ \frac{{\omega_p}^2(\omega+q)}{\omega(\omega^2-{\omega_2}^2)+ q(\omega^2-{\omega_1}^2)},
\end{equation}
in terms of a Nevanlinna-class $q=q(k,\omega)$. The frequencies ${\omega_1}$ and ${\omega_2}$ are defined as respective ratios of the moments $C_{\nu}$:
\begin{eqnarray}\label{57}
{\omega_1}^2&=&C_2/C_0={\omega_p}^2[1-\varepsilon^{-1}(k,0)]^{-1},\nonumber\\
{\omega_2}^2&=&C_4/C_2={\omega_p}^2[1+Q(k)],
\end{eqnarray}
where $\varepsilon^{-1}(k,0)$ can be determined from the classical form ($\hbar\to 0 $) of the FDT (thermal equillibrium) eq. (\ref{54}) and the Kramers-Kronig relation \cite{bibArkhipov2008}:
\begin{equation}\label{Kramer}
Re\varepsilon^{-1}(k,\omega)=1+\frac{1}{\pi}P.V.\int_{-\infty}^{\infty}\frac{Im \varepsilon^{-1}(k,\omega)}{\omega'-\omega}d\omega'
\end{equation}
In this way, we get the following expression :
\begin{equation}\label{65}
Re\varepsilon^{-1}(k,0)=1-2 S_{zz}(k)\frac{{k_{De}}^2}{k^2},
\end{equation}
where $Re\varepsilon^{-1}(k,0)=\varepsilon^{-1}(k,0)=\varepsilon^{-1}(k)$,  $S_{zz}(k)$ is defined by (\ref{Szz}). The function defining the second moment is given by \cite{bibAdam93}: \\
\begin{equation}\label{58}
Q(k)=K(k)+L(k)+H.
\end{equation}
It contains the kinetic contribution for a classical system:
\begin{equation}\label{59}
K(k)=3 (\frac{k}{k_{D}})^2,
\end{equation}
 where ${k_{D}}^2={k_{De}}^2=n_e e^2/\varepsilon_0 k_B T$. The contribution due to electron-ion HGK correlations is in our approach represented by:
 \begin{equation}\label{Hnew}
H=\frac{h_{ei}(r = 0)}{3}=\frac{g_{ei}(r = 0)-1}{3}\simeq -\frac{1}{3}.
\end{equation}
\indent Within the screened HGK model, the $H $ in eq. (\ref{Hnew}) can be approximated by  $-1/3$ because we consider the ion structure, that means that the electron cannot approach the ion at $r=0$ distance. Note that $g_{ei}(r = 0)=0$ is not exactly true, it is just a good approximation. The term $L(k)$ takes into account the electronic correlations, in accordance to the approximation $g_{ei}(r = 0)=0$ we calculated it for the Coulomb potential: 
\begin{equation}\label{61}
L(k)=\frac{1}{2\pi^2 n_e }\int_0^{\infty}p^2 [S_{ee}(p)-1] f(p,k)dp,
\end{equation}
 where \\
  \begin{eqnarray}\label{62}
f\left( p,k\right) =\frac{5}{12}-\frac{p^{2}}{4k^{2}}+\frac{\left(k^{2}-p^{2}\right) ^{2}}{8pk^{3}}\ln \left\vert \frac{p+k}{p-k}\right\vert.  
\end{eqnarray}
 In (\ref{61}) the static structure factor is the one defined in (\ref{53}) with the potentials given in (\ref{42}-\ref{45}). 
The authors of \cite{bibAdam93} suggested to approximate $q(k,\omega)$ by its static value $q(k,0)=\imath h(k)$, connected to the static value $S_{zz}(k,0)$ of the dynamic structure factor through eq. (\ref{54}):
\begin{equation}\label{63}
h(k)=\frac{({\omega_2}^2-{\omega_1}^2){\omega_p}^2}{\pi\beta \phi(k){\omega_1}^4 S_{zz}(k,0)}>0,
\end{equation}
with 
\begin{equation}\label{635}
S_{zz}(k,0)\simeq S^{0}_{zz}(k,0), 
\end{equation}
where $S^{0}_{zz}(k,0)=\frac{n_e}{k}\sqrt{\frac{m}{2\pi k_B T}}$ \cite{bibIschimaruStatPlas} so that the normalized dynamic factor takes the following form:
\begin{eqnarray}\label{64}
\frac{S_{zz}(k,\omega)}{S_{zz}(k,0)} &=& \frac{\beta\hbar }{[1-\exp{(-\beta\hbar\omega)}]}\nonumber\\
&\times &\frac{\omega h^2(k) {{\omega_1}^4}} {\omega^2(\omega^2-{\omega_2}^2)+ h^2(k)(\omega^2-{\omega_1}^2)},
\end{eqnarray}
with the more simplified expressions for $h(k)$:
\begin{equation}\label{650}
h(k)=\frac{\varepsilon_0 \sqrt{2\pi k_B T} k^3{\omega_p}^2({\omega_2}^2-{\omega_1}^2)}{\pi\beta \sqrt{m} n_e e^2{\omega_1}^4},
\end{equation}
$\omega_1(k)$, $\omega_2(k)$:
\begin{eqnarray}\label{570}
{\omega_1}^2&=&C_2/C_0=\frac{{\omega_p}^2 k^2}{2{k_{De}}^2S_{zz}(k)},\nonumber \\
{\omega_2}^2&=&C_4/C_2={\omega_p}^2[1+K(k)+L(k)-\frac{1}{3}],
\end{eqnarray}
\indent In Figures \ref{Fig:206} and \ref{Fig:207} the DSF at a moderate temperature $T=30000 K$ and concentrations $n_e=1.741 \cdot 10^{20}-10^{22} cm^{-3}$ but for the values of  $\Gamma_{ee}$ used in \cite{bibAdam93} are shown. As one can see in Figures \ref{Fig:206}, \ref{Fig:207}, the curves for alkali plasmas are different from those given for the HLPC model \cite{bibAdam93}, where the ion structure was not taken into account. In the case of alkali plasmas the curves split. This can be explained by that fact that alkali ion structure influences the dynamic structure factor. In the Figures the position of the central peaks coincides but its intensity in alkali plasma is more pronounced. Positions of the plasmon peaks are slightly shifted. We observe that the curves shift in the direction of lower values of $k$ compared to the corresponding results of \cite{bibAdam93}. In Fig. \ref{Fig:206} (b) the curves split into three very sharp peaks. Observe that with an increase of number of shell electrons from Li$^+$ to Cs$^+$ at low value of $k$ the intensity of the lines grows while at higher value of $k$ it diminishes. In Fig. \ref{Fig:207} the position of the plasmon peaks shift in the direction of higher absolute value of $\omega $, as compared to those in Fig. \ref{Fig:206}. This discrepancy could be also explained by that fact that the considered parameters are extreme, i. e., either high temperature or density and for $\Gamma_{ee}=0.5$, $r_s=0.4$ the degeneration condition is $n_e\lambda_{ee}=0.335$, while for $\Gamma_{ee}=2$, $r_s=1$ the degeneration condition is $n_e\lambda_{ee}=0.678$. \\

\section{Conclusions}

{The $e-e$ , $e-i$, $i-i$ and charge-charge static structure factors have been calculated for alkali and Be$^{2+}$ plasmas using the method described and discussed by Gregori et al. in \cite{bibGreg2006}. The dynamic structure factors for alkali plasmas have been calculated using the moment approach \cite{bibAdam83}, \cite{bibAdam93}.  The screened Hellmann-Gurskii-Krasko potential, obtained on the basis of Bogolyubov's method, has been used taking into account not only the quantum-mechanical effects but also the ion structure \cite{bib73}. The results obtained within the screened HGK model for the static and dynamic structure factors have been compared with those obtained by Gregori et al. and in \cite{bibAdam93} for the hydrogen-like point charges model. We have found small deviations (in the values of the SSFs) from results obtained by Gregori et al while there are significant differences between DSFs. Nevertheless, we have noticed that the present results are in a reasonable agreement with those of  \cite{bibAdam93} at higher values of $k$ and with increasing $k$ the curves damp while at lower values of $k$ we observe sharp peaks also reported in \cite{bibAdam93}. At lower $\Gamma_{ee}$ the curves for Li$^+$, Na$^+$, K$^+$, Rb$^+$ and $Cs^+$ do not differ while at higher $\Gamma_{ee}$ the curves split. As the number of shell electrons increases from $Li^+$ to $Cs^+$ at low $k$ the intensity of the lines grows, while at higher $k$ it diminishes. The difference is due to the short range forces which we took into account by the HGK model in comparison with the hydrogen-like point charges model. One should also take into account that we employed different plasma parameters because at the parameters used in \cite{bibAdam93} the alkali plasmas with closed shell cannot exist.

\begin{acknowledgement}
 The work has been fulfilled at the Humboldt University at Berlin (Germany). One of the authors (S. P. Sadykova) would like to express sincere thanks to the Erasmus Mundus
Foundation, especially to Mr. M. Parske,  for the financial and moral support, to the Institute of Physics, Humboldt University at Berlin, for the aid which made the participation at Conferences possible; I.M.T. acknowledges the financial support of the Spanish Ministerio de Educaci\'{o}n y Ciencia Project No. ENE2007-67406-C02-02/FTN.
\end{acknowledgement}

%
%
%
%



\end{document}